# Symmetry-Breaking of Turbulent Flow in Periodic Porous Media at Intermediate Porosities

Vishal Srikanth[1], Andrey V Kuznetsov[1,§]

[1]Department of Mechanical and Aerospace Engineering, North Carolina State University, Raleigh, NC 27695

[§]Correspondence author. Email: avkuznet@ncsu.edu

**ABSTRACT** This paper presents a novel discovery of a symmetry-breaking effect in porous media with porosity between 0.8-0.9, which we are referring to as the intermediate porosity flow regime. Using large eddy simulation, we studied how heat transfer and turbulent convection occurs within these materials at a microscopic level. We observed symmetry-breaking in porous structures made of regularly spaced circular cylinders – a common design in heat exchangers – immediately following the laminar to turbulent flow transition between Reynolds numbers of 37 and 100. Asymmetric patterns persisted up to Reynolds numbers of 1,000. The initial breakdown of symmetry occurs through a Hopf bifurcation, creating an oscillating flow pattern as shear layers interact around the solid obstacles. When the flow becomes turbulent, random variations in the timing of vortex oscillations (caused by the secondary instability) create asymmetric distributions of fluid velocity and temperature throughout the porous space. This leads to the formation of alternating channels with high and low velocity fluid flow. At the macroscale level, this loss of symmetry creates residual transverse drag force components and asymmetric heat flux distribution on the solid obstacle surfaces. Interestingly, the oscillating flow pattern promotes attached flow on the circular cylinder surfaces, which enhances heat transfer from the cylinders to the fluid. We observe that this secondary flow instability is the primary mechanism of enhanced turbulent heat flux from porous media with circular cylinders compared to those with square cylinders.

## 1. INTRODUCTION

Transition to turbulence in porous media occurs at low pore-scale Reynolds number values that are of the order of 100 (Seguin et al., 1998). Thus, this type of flow is ubiquitous and important across many applications, from industrial heat exchangers (Nield and Bejan, 2017), cooling systems for electronics (Zhao and Lu, 2002), and forest fire modeling (Mell et al., 2009). The solid obstacles that compose the porous medium in the above flow scenarios are often cylindrical in shape, positioned close enough that vortices forming behind one solid obstacle interact with the next downstream obstacle. A better understanding of the characteristics of turbulent flow in porous media can lead to the development of robust and physically accurate turbulence models. This will enable intelligent engineering of the next-generation porous materials whose microscale geometries produce optimal thermal performance.

Research into periodic porous media has revealed how pore geometry constrains turbulence scales (He et al., 2019; He et al., 2018; Chu et al., 2018; Uth et al., 2016; Nguyen et al., 2019) and generally prevents macroscale turbulent structures (Jin *et al.,* 2015; Jin and Kuznetsov, 2017). While very high porosity materials (>98%) might support macroscale turbulence (Rao and Jin, 2022), this requires further verification in non-periodic simulation domains. However, microscale turbulent flow has been extensively studied in recent literature. Microscale turbulent flow structures are generated from the microscale vortices formed behind the solid obstacles (Srikanth et al., 2021), which are then advected by the flow in the pore space where they are dissipated under the action of shear and pressure gradients. Therefore, the geometry of the pore space and the corresponding flow stress distributions determine the properties of microscale turbulence and the related macroscale flow variables such as drag and heat flux. These observations have been documented for a wide range of geometries and

Reynolds numbers (He et al., 2019; Chu et al., 2018; Srikanth et al., 2021; Huang et al., 2022), with notably a strong influence of the turbulent flow characteristics on the geometry of the porous medium.

In this paper, we categorize turbulent flow in porous media into three porosity-based flow regimes: low, intermediate, and high porosity ($\varphi$) flow regimes. In low porosity materials ($\varphi<0.8$), closely-spaced solid obstacles create narrow interconnected channels where small recirculating vortices form, limited by available space. This regime can exhibit flow symmetry-breaking due to the formation of an adverse pressure region in the pore space (Srikanth et al., 2021). In the present study, we investigate a symmetry-breaking phenomenon that occurs in the intermediate porosity flow regime ($0.8<\varphi<0.95$) for cylindrical solid obstacles with circular cross-section. Here, the diameter of the microscale vortices that are formed behind the circular cylinders is approximately equal to the radius of the cylinder and smaller than the pore size by at least a factor of 2. There is adequate pore space in between the solid obstacles in the intermediate porosity regime to allow the development of shedding vortices. However, the vortices interact with the neighboring solid obstacles at a downstream location resulting in complex flow phenomena and flow instabilities that we discuss in this paper. In high porosity media ($\varphi>0.95$), the ample space between obstacles allows flow patterns around each obstacle to develop independently.

## 2. NUMERICAL METHOD

In this paper, we are investigating turbulent flow inside a periodic porous medium composed of cylindrical solid obstacles (Figure 1). The cylindrical obstacles that have circular and square cross-sections are placed in an in-line arrangement separated by a distance, $s$, which we are calling the pore size. We numerically simulate the turbulent flow inside the pores at the microscale level by using Large Eddy Simulation (LES). We model a Representative Elementary Volume (REV) of the porous medium consisting of 4x4 arrangement of solid obstacles (Dimensions – $4s$ x$4s$ x $2s$). We apply periodic boundary conditions at all of the boundaries of the REV. We sustain flow at a constant pore scale Reynolds number ($Re$) by applying a momentum source term $g_i$. The Reynolds number is calculated using the hydraulic diameter of the solid obstacle, $d$, and the double-averaged (time and space) $x$- velocity, $u_m$. The governing equations of momentum are non-dimensionalized by setting the fluid density, $d$, and $u_m$ equal to 1 and the dynamic viscosity equal to $1/Re$. The governing equation of thermal energy is solved as a passive scalar. The Prandtl number of the fluid is set equal to 7 (similar to water at NTP). We define non-dimensional temperature ($\theta$) as $(T_w-T)/(T_w-T_{in})$, where $T$ is the local temperature, $T_{in}$ is the bulk inlet temperature, and $T_w$ is the wall temperature.

We use ANSYS Fluent 16.0 to solve the governing equations for LES with the Dynamic One-equation Turbulence Kinetic Energy (DOTKE) subgrid scale model by applying the finite volume method. The governing equations and details of the model implementation are available in (ANSYS Inc., 2016) and in our previous work (Huang et al., 2022; Srikanth et al., 2021) along with validation studies for the momentum and thermal transport models and convergence of the grid and REV sizes. We demonstrated that LES with DOTKE subgrid scale modeling reproduces the experimentally measured pressure and Nusselt number distributions on the surface of tubes in an in-line tube bank (Aiba et al., 1982). We have also demonstrated that a grid resolution of $\Delta x/s = 0.02$ is sufficient to resolve a substantial portion of the turbulence kinetic energy (volume-averaged LES index of quality (Celik et al., 2005) = 0.95) until the magnitude of turbulence kinetic energy declines by 3 orders of magnitude at the smallest resolved length scale when compared to the large scale eddies. Stretched grid cells of size $\Delta y/s = 0.001$ are used near the solid obstacle walls to resolve the near-wall boundary layer accurately. The maximum value of $\Delta y^+ = 1.5$ and the grid cell growth rate is 1.2, which places ~3 grid cells in the laminar (linear) sublayer even for the regions with the maximum shear.

We use bounded second-order central differencing discretization for the convective terms and second-order central differencing discretization for the diffusive terms of the momentum equation. We use the QUICK scheme for the convective term of the thermal energy equation. We solve the momentum and pressure Poisson equations in a segregated manner using the PISO algorithm (Issa, 1986). We solve the thermal energy equation at the end of the time step. We advance the simulation in time by using a second-order (implicit) backward Euler method. The simulations span 400 non-dimensional time units (17,000 CPU-Hours on 40 threads of Intel Xeon E5649) to converge the turbulence statistics, which corresponds to 200 flow-through cycles for a unit cell. We have confirmed that the modes of symmetry-breaking reported in the paper are not caused by unconverged turbulence statistics by verifying that the same modes are observed over 2 different time intervals.

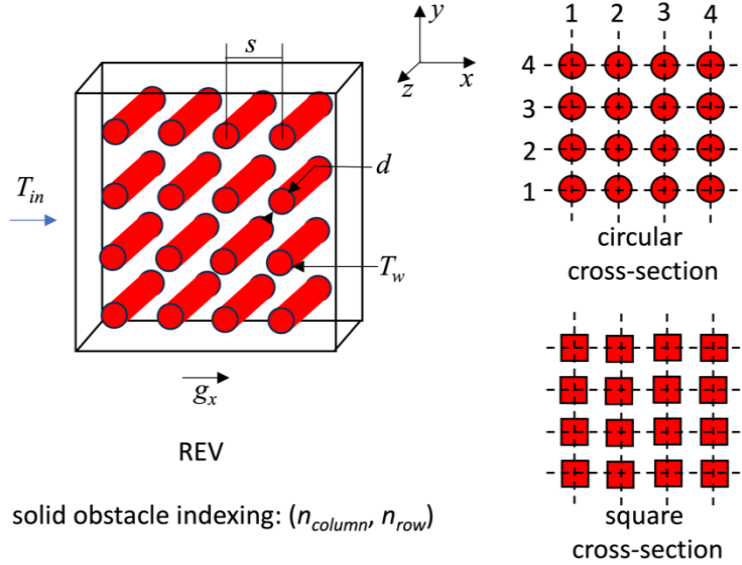

Figure 1. The geometry of the computational domain used to simulate turbulent flow inside the REV of the porous medium.

## 3. RESULTS AND DISCUSSION

Secondary flow instabilities and symmetry-breaking phenomena in porous media are sensitive to two key parameters: porosity and Reynolds number, and symmetry-breaking is often conditional on the smoothness of the solid obstacle surface. A case study of the variation of porosity is used to characterize the symmetry-breaking phenomenon that occurs in the intermediate porosity flow regime. We report the effect that the asymmetrical flow pattern has on turbulent convection heat transfer.

Turbulent flow in porous media breaks symmetry in both the low and intermediate porosity flow regimes. Symmetry-breaking occurs due to the interaction of the flow around a solid obstacle with the neighboring solid obstacle. Therefore, the distance between the solid obstacle surfaces, which is determined by the porosity, is the primary variable that controls the occurrence of symmetry-breaking. Consider turbulent flow at a Reynolds number of 300 inside a periodic porous medium consisting of cylindrical solid obstacles with circular cross-section. In the low porosity flow regime ($\varphi < 0.8$), a secondary flow instability is formed due to competing pressure and inertial force components in the confined pore space, which drives the flow to break symmetry. A pair of shear layers form around the solid obstacle surface starting in the location where the flow separates (Figure 2(a) and (b)). The secondary flow instability at low porosity is characterized by the Kelvin-Helmholtz instability since the pair of shear layers do not interact due to the confined pore space and the von Karman instability cannot be formed (until symmetry is broken). Symmetry-breaking at low porosity manifests above a critical Reynolds number ($Re = 350$-$500$) with deviatory flow streamlines resulting in a change in the direction of the macroscale flow at an angle that is determined by the value of porosity. The symmetry-breaking phenomenon at low porosity is described in detail in our previous work (Srikanth et al., 2021). We noted that deviatory flow terminates at approximately $\varphi = 0.8$ and demarcates the transition between the low and intermediate porosity flow regimes.

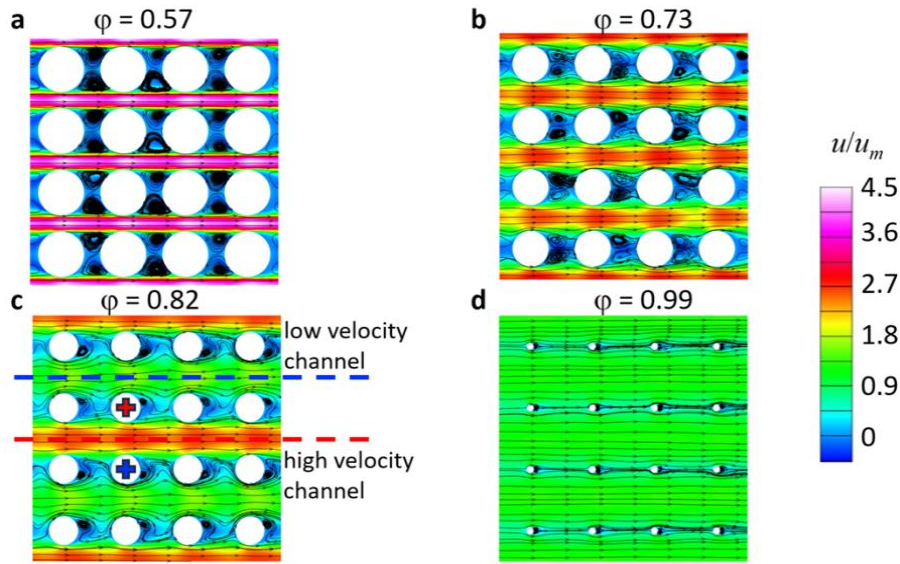

Figure 2. Reynolds-averaged flow streamlines plotted for different values of porosity shows that symmetry-breaking occurs for φ = 0.82 (intermediate porosity) at $Re$ = 300. The colors show the Reynolds-averaged $x$- velocity distribution.

For turbulent flow at $Re$ = 300 in the intermediate porosity flow regime (0.8 < φ < 0.95), the pair of shear layers that are formed around the solid obstacle surface interact with one another in a manner that is similar to von Karman vortex shedding (Figure 2(c)). Interaction between the shear layers is possible at intermediate values of porosity by the presence of an adequate separation distance between the solid obstacle surfaces. However, unlike in the case of external flow around a single isolated solid obstacle, the interaction of the shear layers and subsequent formation of the von Karman vortex street is influenced by the presence of the neighboring solid obstacles downstream of the vortex wake. The result is a newly discovered instability that combines the von Karman vortex shedding and a secondary flow instability. The occurrence of the traditional von Karman vortex shedding behind the solid obstacle is hindered by the possibility that the vortices will immediately impinge on the downstream solid obstacle. A secondary flow instability occurs to divert the path of the shedding vortices such that it circumnavigates the solid obstacle into the pore space without directly impinging on the solid obstacle surface. Thus, the von Karman vortex shedding is superimposed on the oscillations of the secondary flow instability such that vortices are shed into the pore space above and below the solid obstacle in an alternating manner (Figure 3). We note that this secondary flow instability is not observed in the high porosity flow regime due to the large separation distance between the adjacent solid obstacle surfaces (Figure 2(d)).

In the present case where the solid obstacle cross-section is circular, the secondary flow instability introduces an oscillatory shift in the stagnation and separation points on the solid obstacle surface (Figure 3). Consider the path of the vortex shedding behind the solid obstacle on the top right location in Figure 3 (marked with a red plus). In Figure 3(a), the vortex shedding path is tilted in the positive $y$- direction (positive phase) such that the separated flow region on the solid obstacle surface occupies the top-right quadrant. Note that a separation bubble is formed in the bottom-right quadrant, which is followed by flow reattachment on the solid obstacle surface. Correspondingly, the stagnation point is located in the bottom-left quadrant of the solid obstacle surface. When the oscillation caused by the secondary flow instability proceeds in time (Figure 3(b)), the locations of the separated flow region and the stagnation point on the solid obstacle surfaces are mirrored when compared to Figure 3(a) (negative phase). It is evident from the streamline plots in Figure 3 that the oscillatory behavior of the secondary flow instability introduces a shift in the separation and stagnation points along the solid obstacle surfaces and increases the fraction of the solid obstacle surface that is occupied by the attached flow regions when compared to the separated flow regions.

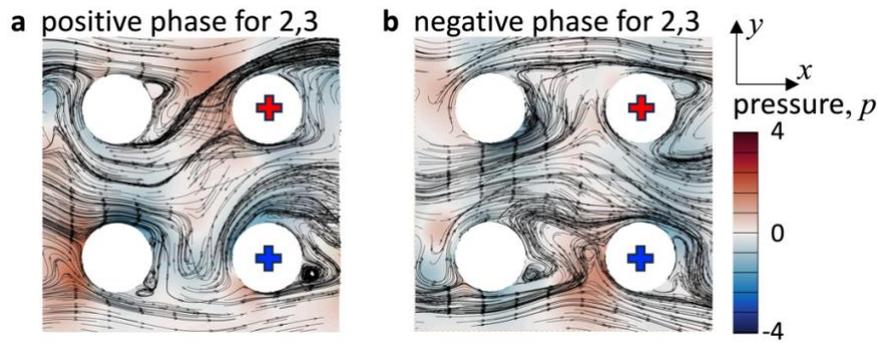

Figure 3. Instantaneous flow streamlines plotted at 2 different time steps (a and b) separated by 28 non-dimensional time units show the oscillations of the secondary flow instability that causes vortex shedding to alternate between the pore space above and below the solid obstacles. The colors show the instantaneous pressure distribution.

When the instantaneous flow distribution is Reynolds-averaged, the Reynolds-averaged velocity (Figure 2(c)) and pressure distributions exhibit a lack of symmetry. The Reynolds-averaged velocity distribution consists of alternating high and low velocity channels in the transverse direction that is perpendicular to the direction of the applied pressure gradient ($g_x$). Here, the term channel refers to the pore space in between the solid obstacles oriented along the direction of the flow (red and blue dashed lines in Figure 2(c) correspond to high and low velocity channels). Close examination of the Reynolds-averaged flow streamlines in Figure 2(c) also reveal asymmetrical recirculation regions as well as an offset in the stagnation point with respect to the geometric plane of symmetry. In the present flow setup, the geometric plane of symmetry that we are referring to is one that bisects the solid obstacle cross-section and is oriented parallel to the *xz*- plane. This type of symmetry-breaking occurs only in the intermediate porosity flow regime when the oscillatory vortex shedding path emerges due to the secondary flow instability.

These high and low velocity channels can also be observed in the simulated mean streamwise velocity profile across a porous medium composed of circular tubes (Kim et al., 2023). The simulation setup of Kim *et al.* (2023) is different when compared to the simulation domain used in the present work. Kim *et al.* (2023) consider a partially porous channel geometry, whereas the present work considers a fully periodic domain. We also note that the porosity of the porous medium used in Kim *et al.* (2023) is 0.75, which is in the low porosity flow regime as per the definition used in the present work. However, we note that a porosity of 0.75 is close to the boundary between the low and intermediate porosity flow regimes (separated at $\varphi = 0.8$). The velocity profile measurements for $\varphi = 0.75$ (Kim et al., 2023) reveal that the centerline velocity in the pore space changes by 15.1% from the low velocity channel to the high velocity channel (Figure 4(a)), indicating that it may also be feasible to observe symmetry-breaking experimentally. Similarly, the velocity profile simulated in the present work for $\varphi = 0.82$ has a 25.3% change in the centerline velocity from the low velocity channel to the high velocity channel (Figure 4(b)).

Note that the magnitudes of the dimensionless *x*- velocity are different for the present simulation and in Kim *et al.* (2023) such that the dimensionless *x*- velocity observed in the periodic porous medium is almost twice the value of observed dimensionless *x*- velocity observed in the velocity channel. For the periodic porous media, the *x*- velocity is non-dimensionalized with respect to the double-averaged (space and time) *x*- velocity in the periodic REV. For the partially porous channel, the *x*- velocity is non-dimensionalized with respect to the inlet velocity of the wind tunnel section. Additionally, only a portion of the wind tunnel width is occupied by solid obstacles and the remaining portion is a clear fluid region. With the available information about the simulation setup and velocity distribution, it was not possible to non-dimensionalize the present simulation with respect to the same characteristic velocity as Kim *et al.* (2023). Therefore, we only present a qualitative comparison of the formation of high and low velocity channels due to the asymmetrical flow distribution.

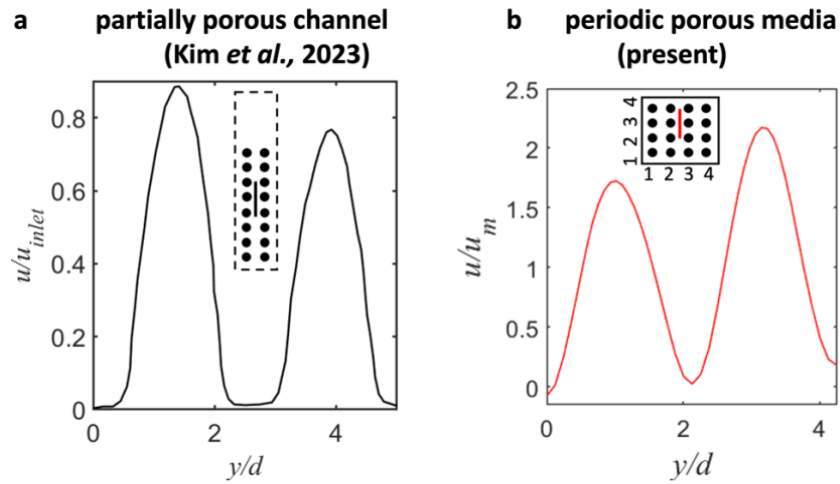

Figure 4. *x*- velocity profiles (a) partially porous media in Kim *et al.* (2023) and (b) periodic porous media in the present work in the direction perpendicular to the streamwise direction. The velocity profile shows the formation of high and low velocity channels in the pore space caused by the symmetry-breaking phenomenon. The black and red lines in the sketches show the location of the velocity profile.

Unlike the deviatory flow symmetry-breaking that occurs at low porosity, symmetry-breaking at intermediate porosity is not apparent in the instantaneous flow distribution. The high and low velocity cells develop inside the pores when the flow is averaged over time due to asymmetrical oscillations of the vortex shedding path caused by the secondary flow instability. Low velocity cells are formed inside the pores where the vortex shedding duration is longer, whereas high velocity cells are formed in the pores where the vortex shedding duration is shorter. The vortex shedding duration is visualized in the present work by calculating the *y*- pressure force that is acting on the surface of the solid obstacle (Figure 5). If the flow is symmetric, the Reynolds average of the *y*- pressure drag will be equal to zero. In the present case, there exists a bias in the *y*- pressure drag that acts in either the positive or negative *y*- direction depending on the offset location of the stagnation point. Positive bias in the *y*- pressure drag is caused because the duration of the positive phase of the vortex shedding process is greater than the duration of the negative phase (solid obstacle (2,3) in Figure 5). The asymmetry in the vortex shedding over time is also inferred from the shape of the plot of *y*- pressure drag in Figure 5. Positive bias in the *y*- pressure drag for the solid obstacle results in a low velocity cell in the pores above the solid obstacle. The vice-versa is observed for the solid obstacles with negative bias in the *y*- pressure drag.

We note that the shift in the separation point and the oscillation of the separated flow region on the solid obstacle surface is less intense for cylindrical solid obstacles with square cross-section even though the porosity is the same as the case with circular solid obstacle cross-section. The vertices of the square geometry prescribe the location of the separation points and limit the oscillation of the vortex shedding path. As a result, vortices shed behind square solid obstacles impinge on the adjacent solid obstacles that are downstream, unlike for circular solid obstacles. The size of the vortex structures as well as the fraction of the solid obstacle surface area occupied by separated flow is larger for square solid obstacles when compared to the circular solid obstacles.

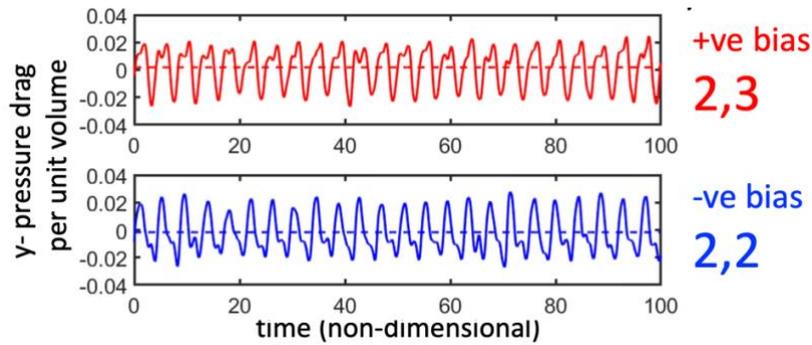

Figure 5. The time series of *y*- pressure drag force acting on the solid obstacle surface showing the bias in the duration of the vortex shedding phases that ultimately results in asymmetrical Reynolds-averaged velocity and pressure distributions. The dashed lines show the Reynolds-averaged *y*-pressure drag value for the solid obstacle. The locations of the solid obstacles (2,2 and 2,3) are shown in Figure 1.

Symmetry-breaking for flow around cylindrical solid obstacles with circular cross-section is characterized by the occurrence of the secondary flow instability and an oscillatory vortex wake path resulting in attached flow on the solid obstacle surface. These conditions are beneficial to increase the heat flux from the solid obstacle surface to the surrounding fluid. Attached flow on the solid obstacle surface decreases the thermal boundary layer thickness when compared to separated flow, which in turn increases the heat flux from the solid obstacle surface in that region. Additionally, smaller vortices are formed behind the solid obstacles with circular cross-section due to the shift in the separation point caused by the secondary flow instability. There are two factors that cause the improvement of heat transfer from the solid obstacle surface in porous media: (1) small size of the microscale vortices results in a smaller fraction of the solid obstacle surface area covered by recirculating flow and increases local heat flux, and (2) frequent vortex shedding from the solid obstacle surface advects recirculating vortex structures with high core temperature and increases local heat flux. The flow around solid obstacles with circular cross-section combines both of these factors and results in higher surface-averaged heat flux when compared to the solid obstacles with square cross-section. This is further supported by the occurrence of a lower vortex core temperature for solid obstacles with circular cross-section when compared to square cross-section (Figure 6(a)). The core temperature of the vortex increases due to recirculation, but the vortex is subsequently advected during the rapid vortex shedding process. Therefore, the secondary flow instability that is observed for flow around solid obstacles with circular cross-section promotes heat transfer in porous media in the intermediate porosity regime. When the heat transfer rate is compared for solid obstacles with circular and square cross-section (at $\varphi = 0.87$, $Re = 300$), the circular solid obstacles have 5% greater double-averaged heat flux (and 7% lesser coefficient of drag) when compared to square solid obstacles.

When the temperature distribution is Reynolds-averaged in the case of solid obstacles with circular cross-section, symmetry-breaking in the velocity distribution causes asymmetry in the Reynolds-averaged temperature distribution (Figure 6(b)). Consequently, the distribution of the Reynolds-averaged heat flux on the solid obstacle surface is asymmetrical (Figure 6(c)). The heat flux has a higher magnitude on the surface of the solid obstacle that is exposed to the high velocity channel when compared to the low velocity channel. However, this asymmetry does not appear to have any effect on the double-averaged Nusselt number over the entire REV since the increase in the solid obstacle surface heat flux caused by the high velocity channel is counterbalanced by the decrease in the heat flux caused by the low velocity channel.

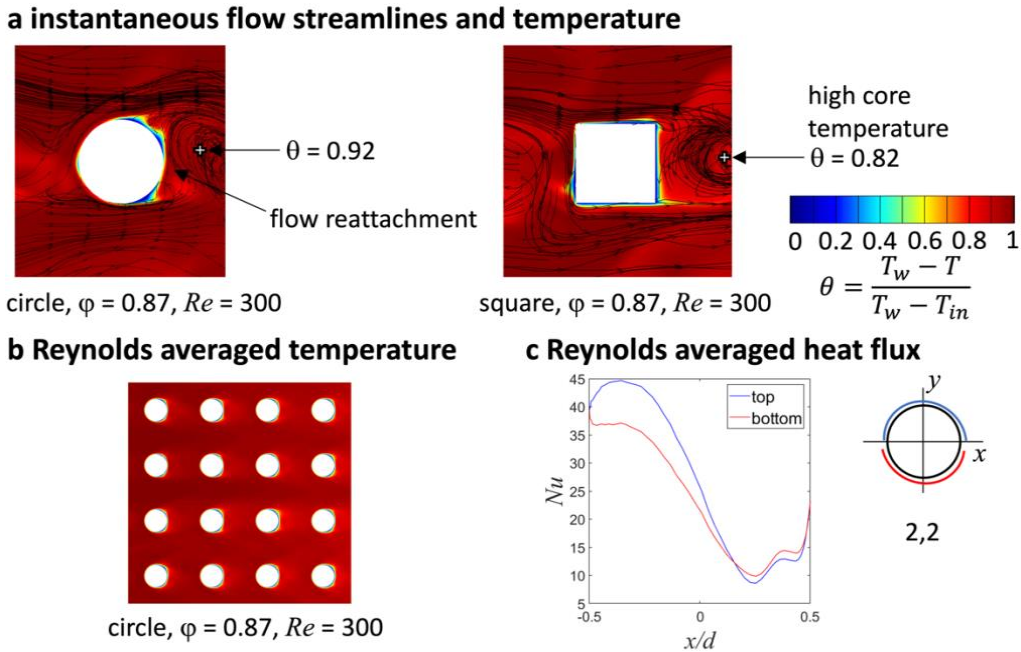

Figure 6. (a) the instantaneous temperature distribution around solid obstacle (2,2) reveals that vortex structures are smaller and have lower core temperature for circular cross-section when compared to square cross-section. (b) The effect of symmetry breaking on the Reynolds-averaged temperature distribution is not as prominent as the velocity distribution. (c) Symmetry-breaking causes asymmetric Nusselt number distribution on the solid obstacle surface.

## 4. SUMMARY

In this paper, we have identified three distinct regimes of turbulent flow in porous media based on medium porosity: low, intermediate, and high. These classifications stem from how microscale vortices behave at different porosities, with regime boundaries expected to shift toward higher porosity values as Reynolds numbers increase. Our research focused on turbulent convection heat transfer in the intermediate porosity regime ($0.8<\varphi<0.95$) for porous media composed of cylindrical solid obstacles, such as typically found in heat exchangers. In this regime, we discovered new secondary flow instabilities and symmetry-breaking effects that significantly impact both microscale velocity distribution and heat transfer from the obstacles to the surrounding flow.

The secondary flow instability emerges from interacting shear layers formed around the obstacles during von Karman instability formation. Since the microscale vortices formed behind the solid obstacles are proximal to their downstream neighbor, confined geometry causes a secondary flow instability where the vortex wake behind the solid obstacle starts oscillating around the downstream solid obstacle. These vortex structures are advected into the pore space rather than impinging on the downstream solid obstacle surface. The resulting flow pattern promotes attached flow on the solid obstacle surface due to a shift in the separation point that is caused by the vortex wake oscillations. Notably, this oscillatory vortex wake does not occur for cylindrical solid obstacles with square cross-section since their sharp corners fix the separation points. Attached flow on the solid obstacle surface and small vortices caused by the secondary flow instability result in greater double-averaged heat flux on the solid obstacle surface and lower vortex core temperature for solid obstacles with circular cross-section when compared to square cross-section.

These wake oscillations show temporal asymmetry - one phase of the oscillation spans a greater duration than its pair. When averaged over time, this creates asymmetric Reynolds-averaged velocity and temperature distributions, with the formation of pronounced high and low velocity channels in the pore space. This is because the temporal asymmetry in vortex shedding causes vortices to be preferentially advected into a particular pore space. When vortices concentrate along a single pore region, a low velocity channel is formed over time. Consequently, the heat flux distribution also becomes asymmetric where the solid obstacle surface exposed to the high velocity channel experiences higher heat flux from the solid obstacle surface to the surrounding fluid. However, the increase in the heat flux caused by the high velocity channel is counterbalanced by the decrease in the heat flux by the low velocity channel. However, the enhanced heat transfer performance of circular obstacles compared to square ones primarily stems from the secondary instability itself, not from this asymmetric heat flux distribution, which is merely a byproduct of the oscillating wake pattern.

## 5. ACKNOWLEDGEMENTS

The authors acknowledge the computing resources provided by North Carolina State University High Performance Computing Services Core Facility (RRID:SCR_022168). This research was funded by the National Science Foundation award CBET-2042834.

## 6. NOMENCLATURE

| Quantity | Symbol | Units |
|---|---|---|
| Porosity | $\varphi$ | - |
| Pore size | $s$ | - |
| Hydraulic diameter of the solid obstacle | $d$ | - |
| Applied pressure gradient | $g_i$ | - |
| Double-averaged $x$- velocity | $u_m$ | - |
| Maximum grid cell size | $\Delta x$ | m |
| Near-wall grid cell size | $\Delta y$ | m |
| Near-wall grid cell size in wall coordinates | $\Delta y^+$ | - |